# Inferring High Quality Co-Travel Networks

Youfang Lin[1,3*], Xuguang Jia[2], Mingjie Lin[1], Steve Gregory[3], Huaiyu Wan[1,4], Zhihao Wu[1]

**1** School of Computer and Information Technology, Beijing Jiaotong University, Beijing, China, **2** TravelSky Technology Limited, Beijing, China, **3** Department of Computer Science, University of Bristol, Bristol, UK, **4** Department of Computer Science and Technology, Tsinghua University, Beijing, China

**Abstract**

Social networks provide a new perspective for enterprises to better understand their customers and have attracted substantial attention in industry. However, inferring high quality customer social networks is a great challenge while there are no explicit customer relations in many traditional OLTP environments. In this paper, we study this issue in the field of passenger transport and introduce a new member to the family of social networks, which is named Co-Travel Networks, consisting of passengers connected by their co-travel behaviors. We propose a novel method to infer high quality co-travel networks of civil aviation passengers from their co-booking behaviors derived from the PNRs (Passenger Naming Records). In our method, to accurately evaluate the strength of ties, we present a measure of Co-Journey Times to count the co-travel times of complete journeys between passengers. We infer a high quality co-travel network based on a large encrypted PNR dataset and conduct a series of network analyses on it. The experimental results show the effectiveness of our inferring method, as well as some special characteristics of co-travel networks, such as the sparsity and high aggregation, compared with other kinds of social networks. It can be expected that such co-travel networks will greatly help the industry to better understand their passengers so as to improve their services. More importantly, we contribute a special kind of social networks with high strength of ties generated from very close and high cost travel behaviors, for further scientific researches on human travel behaviors, group travel patterns, high-end travel market evolution, etc., from the perspective of social networks.

Funding: This work was supported by research support programme of Civil Aviation Administration of China (CAAC), project number: MHRD201130. This work was also supported by the Fundamental Research Funds for the Central Universities of China.
* E-mail: yflin@bjtu.edu.cn

## 1. Introduction

A social network consists of a set of persons or groups each of which has one or more types of relations with others. Social networks provide us a new perspective to investigate the characteristics and patterns of human social behaviors. Social network analysis [1, 2] has attracted more and more interest in the past decade.

In many traditional OLTP environments, there may exist millions of, or even more, customers while no explicit customer relationships exist in these environments. Therefore, inferring customer social networks is the most basic and often challenging task for analyzing customers from the perspective of social networks. Generally, it is impossible for us to construct a completely "true" social network, due to the complexity and unavailability of data. The common practice is to reconstruct partial social networks in different fields by discovering explicit or latent connections between actors based on the data which records the interactive behaviors among people. For examples, communication social networks are reconstructed based on the records of communication events (e.g., emails, phone calls, instant messages, etc.)

among social members; online social networks are constructed from various online interactions in social network sites (e.g., Facebook, Twitter, LinkedIn, etc.); affiliation social networks are inferred from the facts that a group of individuals attend common activities (e.g., collaborating works, joining the same organizations, etc.).

All these kinds of social networks can be used for general social network analyses, e.g., statistical analysis [3, 4], community detection [5], link prediction [6], evolution analysis 7], social influence analysis [8], information diffusion [9], etc. Meanwhile, social networks from different fields can also be used for their special mining tasks. For example, co-author and citation networks can be used for expert finding [8] and paper recommendation [10]; email networks can be used for anti-spam email filtering [11]; online social networks can be used for public opinion analysis [12]; etc.

In this paper, we add a new member to the family of social networks, which is named co-travel networks. We propose this new concept to represent the social networks which may be constructed based on the travel records accumulated in passenger transportation systems, such as civil aviation or railway systems, and reflecting the travel behaviors of real social members. We assume that any two people who travel together should have some kind of social relationships such as family, colleague, friendship, etc. A co-travel network is a type of affiliation network, in which a group of individuals who join the same travel are linked with each other. It should be noted that co-travel means not only flying or riding together, but also having a common journey.

We study how to construct high quality co-travel networks and propose a feasible solution to infer ties between passengers in a passenger transportation system which provides group ticketing services. We test our ideas on an encrypted two-year civil aviation dataset and a large-scale, high quality passenger social network (i.e., co-travel network) is successfully constructed.

Compared to other kinds of social networks, the high quality of our co-travel networks is mainly reflected in the following two aspects:

- ***The absolute authenticity of the individuals and the ties***. The existence of all kinds of ambiguities of the individuals and the ties in data sources is a very important factor that may influence the quality of social networks. For example, one person may have multiple accounts(so called sock puppets) in an online social network; large numbers of spam mails and bulk mails suffuse email networks; large numbers of duplicate names exist in co-author networks; harassing calls and wrong calls flood mobile social networks; etc. However, in our co-travel networks, every passenger is unique since he/she has an unique ID in the transportation system. Meanwhile, the ties between passengers in our co-travel networks have almost no noise. As we know, in the real world, if one person travels together with another people, it means that there is a strong social relationship between them most of the time, and this ensures the authenticity of the ties. There is only one exception is that, in large tour groups organized by travel agencies, the members have a common journey but may not be acquainted with each other, and this can be a kind of noise in co-travel networks. In this paper, we propose a feasible noise processing strategy to handle such temporary co-travel ties and make our co-travel networks authentic.

- ***The high strength of the ties.*** The strength of ties is a very important property of a social network. Here, we consider the strength of ties of social networks from an overall perspective. By Granovetter's definition [13], the strength of a tie is a combination of the amount of time, the emotional intensity, the intimacy (mutual confiding), and the reciprocal services which characterize the tie. Researchers have proposed many different measures to quantify the strength of ties in a given social network context, most of which are based on the frequency or duration of attending the reciprocal or group activities. Obviously, different types of activities cost attendees differently and have different geographic distance among them. It is fairly understandable that attending a higher cost or geographically closer activity indicates a higher strength of ties overall. For instance, we can say that the overall strength of ties of communication networks are much weaker than that of co-author networks, because collaborating on a paper needs a much higher cost than writing an email or make a phone call and requires the co-authors to work together in the same place for a long time in most cases. According to this idea, the strength of ties in co-travel networks is relatively high, since travelling together is a kind of face-to-face and high-cost activity.

For understanding the characteristics of co-travel networks, we study a range of statistics on a large-scale network which contains more than 75 million distinct passengers and more than 320 million distinct co-travel relationships. The nationwide coverage and structural diversity of the social members in the network make our analyses more general and representative. These statistics indicates that co-travel networks take on many different properties from other kinds of social networks, for example their sparsity and high aggregation (see Section Results for details).

In addition to general social network analyses, co-travel networks have a wide range of applications in the passenger transportation domain. For examples, we can analyze the passenger behaviors from the network perspective and some interesting travel patterns can be found; typed community structures (e.g., families, organizations, tour groups) and special roles (e.g., tour guides, leaders) can be easily discovered; the commercial values of customers can be evaluated more accurately; the evolving analysis on co-travel networks can be used for understanding the laws of passenger transportation market; etc. All these analyses can provide great decision supports for passenger transportation service and planning.

The rest of the paper is organized as follows. Section Related Works gives an overview of the related works. In subsection Civil Aviation Co-Travel Networks, we introduce related concepts of co-travel networks and the background of civil aviation. In subsection Inferring High Quality Co-Travel Networks, we describe our method for inferring high quality co-travel networks from civil aviation passenger travel records. Then in Section Results we study the characteristics of co-travel networks by investigating a range of statistics. Next, in Section Discussion, we simply present some potential applications of co-travel networks. Finally in Section Conclusion, we conclude this paper and discuss some directions for future work.

## 2. Related Works

As we know, "real" social ties cannot be directly observed and hence must be inferred from the data records of all kinds of interactions or communications among social members. Researchers have inferred many kinds of social networks in different fields. For example, email social networks are constructed from real email exchange records [14, 15, 16]; mobile social networks are constructed from phone call records between mobile users [17, 18, 19, 20]; online social networks collected from all kinds of online interactions [21, 22];affiliation networks are constructed based on the activities of collaborating scientific papers [3, 4], being actors or directors in the same films [23], joining the same terrorist organizations [24], etc.

In the civil aviation domain, Farrugia et al. [25] proposed five rules to infer ties between passengers to enhance airline customer relationship management data. They mainly used the data extracted from Passenger Name Records (PNR) and Travelled-together-X-times to infer ties between passengers but did not consider any noise that may exist in data, especially the false ties collected from large travel groups and random booking records. They used co-appearance times in the same plane to measure the strength of ties, while in some circumstances this may not reasonable. For example, two strangers in a larger travel group and whose journey consists of three flight segments should have a weaker tie than two acquaintances who co-appear only once in a small group. In addition, the authors did not verify the authenticity of the ties between passengers inferred from the travel or booking records.

In [25], the authors also used information of having the same email suffix, mail addresser phone number to infer different kinds of ties between customers. However, different ties inferred from different data may have very different levels of strength. Co-travel in a group is generally an evidence of a very close social relationship, while having the same addresses does not take on the same potency and it can be only used as an extra evidence of the existence of a social relationship. Meanwhile, these kinds of information may not available or complete in many real systems. In this paper, we do not consider using extra customer profiles but focus on using passenger travel behaviors to infer high quality co-travel networks.

With regard to the statistics of social networks, Newman [3, 4] studied the mean papers per author, mean authors per paper, collaborators per author, size of components, clustering coefficient, shortest path, centrality and distance in scientific collaboration networks; Choudhury et al. [15] studied the node degree, neighbor degree, embeddedness, clustering coefficient, network constraint, ego components, fractional network size and network components in communication (i.e., email) social networks. In this paper, we will give a study of these statistics in co-travel networks and try to find out their different characteristics compared with other kinds of social networks.

## 3. Civil Aviation Co-Travel Networks

In this section, we will first introduce some related concepts and notations about co-travel networks, and then present the dataset and its basic statistics.

### 3.1 Definitions

**Def. 1.** A **Co-Travel Network** is a graph $G = (V, E)$, where $V$ is a set of travelers or passengers and $E$ is a set of edges between the co-travelers. If the travelers in $V$ are all civil aviation passengers, we call $G$ a **Civil Aviation Co-Travel Network**.

The edge set $E$ is generally inferred from observed co-travel behaviors. A group travelling by air or by train requires the group members to pay a high cost to co-travel in a close manner. Therefore, travelling in a group, especially in a small group, can be seen as strong evidence of the existence of certain kinds of intimate relationships between group members. And consequently, this kind of tie has a high strength based on the facts of high cost and closeness of the co-travel behaviors.

There are many challenges in constructing such a kind of co-travel social network. The first challenge is how to identify a co-travel passenger group, i.e., which kinds of behaviors can be treated as co-travel events. The second challenge is how to measure the strength of ties in the network. The third challenge is how to filter or utilize the noise that may exist in different passenger group behaviors.

### 3.2 PNR and Passenger Group

In the Airline industry, a Passenger Name Record (PNR) is a record that contains the itinerary for a passenger or a group of passengers travelling together. In the civil aviation industry, PNRs can be seen as an itinerary or a journey record which contains several travel segments identified by flight numbers. The co-appearance of passengers in the same PNR shows that they may have co-booked the tickets together or somebody else (such as tour guides) may have made a group booking for them. Therefore, it is natural to identify the passengers in the same PNR as a passenger group.

If data are available, it may also be possible to identify passenger groups or to strengthen ties according to other group behaviors, such as group check-in, neighboring seating, neighboring boarding, etc. However, the PNR data are the most direct and effective evidence for identifying passenger groups in the civil aviation business process.

We can formally express a PNR as a relation set of passenger groups to flights.

**Def. 2.** We define a $PNR = \{SFPG\}$, where an $SFPG = (V', f)$ is a Single Flight Passenger Group, in which $V' \in V$ is a passenger set, $f$ denotes a flight and the pair $(V', f)$ denotes the passengers $V'$ who book together on the flight $f$.

The simplest method to infer ties between passengers $u$ and $v$ is to find out if they have appeared in the same passenger group(s).

Co-flight times can also be possibly used to infer ties between passengers. We can generate $C(N, 2) = N(N-1)/2$ links between passengers for each flight, where $N$ is the number of passengers on a flight, then merge all the links in the data and establish a tie between any two passengers if there are more than $t$ links between them, where $t$ is a threshold. However, this method will cause too much noise, especially for networks generated with a small $t$, because real ties may be discarded and false ties may be introduced. Therefore, [25] established a Travelled-together-4-times network to eliminate the noise, but many 3 or less times real co-travel ties in small groups would be neglected.

### 3.3 Dataset and Basic PNR and SFPG Statistics

The dataset we used is an encrypted dataset having no private passenger information, under the research support programme of Civil Aviation Administration of China (CAAC). The dataset contains two years Chinese nationwide domestic PNRs, within which flight numbers, places of departure and destination are well encrypted, and passengers are identified only as sequential integral IDs. In this two years dataset, there are approximately a hundred million distinct passengers in more than 5 million flights.

The total number of PNRs is about 256 million, of which 74.46% have only one distinct passenger and the remaining 25.54% contain at least two distinct passengers. However, these 25.54% PNRs contain more than 75 million distinct passengers, and this gives us a very strong indication that a large-scale authentic co-travel social network can be constructed.

Figure 1 and Figure 2 present the frequency distributions of the PNRs and SFPGs with different numbers of distinct passengers respectively. It can be seen that these two distributions are all typical heavy tail distributions. And in these Figures, we also observe a very interesting phenomenon: the curves decrease in a stepped down manner in their middle parts, like a ladder, and in the range of 6 to 40 in the horizontal axes, each tier has a length of 5 and a raised tail. It shows that people tend to take the multiples of 5 (e.g., 10, 15, 20, 25, 30, 35) as a group size while planning large travel groups in real world, and we may call it a **5-multiple preference**.

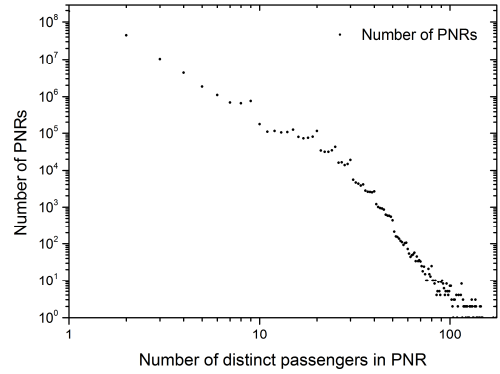

**Figure 1. Frequency distribution of the PNRs with different numbers of distinct passengers**

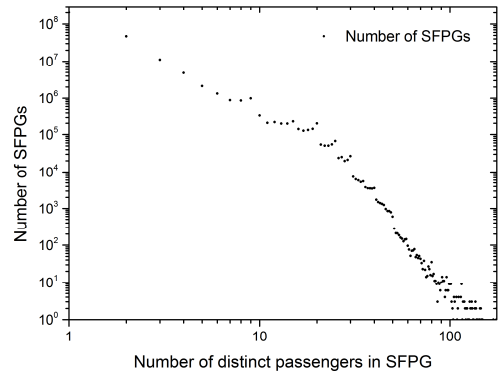

**Figure 2. Frequency distribution of the SFPGs with different numbers of distinct passengers**

Figure 3 presents the frequency distribution of PNRs with different number of flight segments. In the Figure, we can find a very interesting result that the curves are convex at the even numbers of flight segments but concave at the odd numbers of flight segments, and this phenomenon is more obvious for the PNRs with large numbers of distinct passengers. The phenomenon reflects the fact that travelers tend to make round trips rather than one-way trips to a large extent, and it appears that the larger a group is, the higher probability it has to book the return tickets.

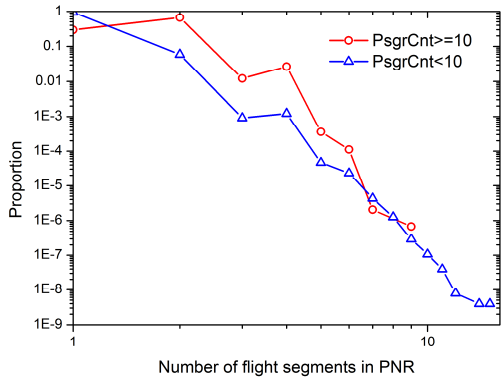

**Figure 3. Frequency distribution of PNRs with different number of flight segments** (PsgrCnt denotes the number of distinct passengers in a PNR and 10 is the business threshold of the size of a large travel group.)

## 4. Inferring High Quality Co-Travel Networks

In this section, we will first introduce the concepts of active ties and passive ties in co-travel networks and the measures of tie strength, and then describe in detail our model for calculating the strength of ties to infer high quality co-travel networks. Finally the definitions of threshold co-travel networks are presented.

### 4.1 Active Ties and Passive Ties

#### 4.1.1 Concepts

Having an insight into the business scenarios of civil aviation, we find that the generation mechanism of co-travel ties is very different from those of ties in other types of social networks. For examples, ties in email networks or scientific collaboration networks are generated by active behaviors (i.e., sending and receiving emails or co-writing publications) between network members. However, co-travel ties could be generated in active or passive ways: in many scenarios, acquainted passengers (e.g., family members, close friends or colleagues) booking tickets together are naturally in the same passenger groups; while in some other scenarios, strangers might be accidentally or intentionally placed into the same passenger group by travel agencies or other organizers. From this viewpoint, we can divide co-travel ties into two categories:

- *Active Ties*, which refers to the ties between passengers who have active co-travel behaviors.
- *Passive Ties*, which refers to the ties between passengers who are passively placed into the same passenger groups.

#### 4.1.2 Business Scenarios of Passive Ties

According to the above concepts, we can simply take the ties between acquaintances as active ties while those between strangers as passive ties. Based on our in-depth investigation into the business of civil aviation, we find several common scenarios in which passive ties may exist:

1) A guide escorts a large tour group in which most of the members are strangers while some small subgroups maybe acquainted with each other.

2) A ticket agency intentionally creates an artificial large group for business discounts, by scraping up strange passengers who book the same flight.

3) An activity organizer books and/or pays the air tickets for a group of strangers who attend the activity and have the same journey.

Actually, if we can distinguish the passive ties from the active ties, they might be very useful in some real applications (e.g., tourism market analysis, group travel pattern analysis, intentional grouping fraud detection, etc.). However, if the passive ties are treated in the same way as the active ties, it is bound to have a negative effect on the quality and further utilization of co-travel networks. Consequently, the key problem is how to determine whether a specific tie is an active tie or a passive tie.

Frankly speaking, it is very hard for us to determine whether any two co-travel passengers are strangers or acquaintances. However, according to the above scenarios and the civil aviation business principles (e.g., a passenger group containing 10 or more persons can be treated as a **Large Passenger Group** (**LPG**) which is able to apply for a group discount price of tickets in China), we find that co-travel strangers have a much larger chance to appear in an LPG than in a **Small Passenger Group** (**SPG**), i.e., LPGs are likely to produce passive ties in general. After introducing the measures and the calculation of tie strength, we will present how to label passive ties and active ties in subsection Labeling Passive Ties and Active Ties.

### 4.2 Measures of Tie Strength

Some simple measures directly summed from detail travel records, e.g., co-flight costs, Co-SFPG or Co-PNR times, can be used to evaluate the strength of ties in co-travel networks. However, these simple measures are not accurate enough to indicate the closeness of ties between passengers in real world. In this paper, we propose a new measure, called co-journey times, to evaluate the strength of ties.

#### 4.2.1 Simple Measures

Firstly, we introduce three simple measures as the strength of ties: Co-Flight costs, Co-SFPG and Co-PNR times.

- **Co-Flight Costs**

A co-flight cost means the common cost of all the co-flights between two passengers in a period of time, which can be simply calculated from the price, duration, or distance of flights. It seems that, for any two passengers, the higher their co-flight cost is, the closer relationship they may have. However, flight costs are highly related with the travel destinations, while determining the destination for a travel can be seen as a random event, so it is hard to say the co-flights with different costs must make a difference to the strength of ties between passengers. From this viewpoint, co-flight costs are not very appropriate measures for the strength of ties in co-travel networks.

- **Co-SFPG and Co-PNR Times**

Co-SFPG and Co-PNR times may be the simplest but effective measures that can be used to describe the strength of ties in co-travel networks. The definitions of SFPG and PNR in subsection PNR and Passenger Group imply that, for any two passengers in the same passenger group, there are two levels of evidence, i.e., their co-appearance in the same SFPG (Co-SFPG) and in the same PNR (Co-PNR), which can be used as the evidence of their co-travel behaviors.

Given a passenger pair $(u, v)$, we use $Co\text{-}SFPG(u, v, i)$ to denote that $u$ and $v$ appear together in the same SFPG $i$ and $Co\text{-}PNR(u, v, j)$ that $u$ and $v$ appear together in the same PNR $j$ (i.e., they appear together in any one SFPG included in PNR $j$). Correspondingly, we respectively use $\#Co\text{-}SFPG_{uv}$ and $\#Co\text{-}PNR_{uv}$ to record the Co-SFPG and Co-PNR times between $u$ and $v$ during a given period of time, and apparently we have $\#Co\text{-}PNR_{uv} \leq \#Co\text{-}SFPG_{uv}$. Naturally, we can use $\#Co\text{-}SFPG_{uv}$ or $\#Co\text{-}PNR_{uv}$ to measure the strength of ties between passengers.

Because a PNR may comprise several SFPGs, as shown in Figure 3, it ought to be more accurate to use a PNR to represent a travel than to use an SFPG, so it is better to use $\#Co\text{-}PNR_{uv}$ to measure the strength of ties rather than $\#Co\text{-}SFPG_{uv}$, which is verified in our experimental results in section Results.

In our experiments, to count $\#Co\text{-}SFPG_{uv}$ and $\#Co\text{-}PNR_{uv}$ while inferring co-travel networks from civil aviation passenger behavior data, we iteratively parse all SFPGs in chronological order and extract the Co-SFPG and Co-PNR events from travel records according to the following procedure:

1) Initialize $\#Co\text{-}SFPG_{uv}$ and $\#Co\text{-}PNR_{uv}$ both to be 0;

2) When we observe an event $Co\text{-}SFPG(u, v, i)$, we increase $\#Co\text{-}SFPG_{uv}$ by 1; And if SFPG $i$ is included in a new PNR $j$ (i.e., the first time the event $Co\text{-}PNR(u, v, j)$ is observed), we increase $\#Co\text{-}PNR_{uv}$ by 1.

#### 4.2.2 A New Measure: Co-Journey Times

In real life, a travel or a journey consists of a series of behaviors of social members in a period of time. These behaviors may include planning schedule, booking transportation tickets, reserving hotel, moving along transportation segments, etc. The data from different aviation operation support systems can only reflect the behaviors that can be observed by these systems, such as booking tickets, check-in, and boarding. From the perspective of transportation, a journey consists of a series of travel segments from the departure to the destination which may include several flight segments operated by different airlines and other kinds of transportation segments such as subway, railway, bus, or taxi.

Consequently, a co-booking record just indicates that a group of passengers have taken one or more flight segment(s) together but not necessarily the whole journey. Neither can a one-segment co-flight behavior adequately represent a whole journey in many circumstances. Furthermore, according to business rules, generally a PNR can only contain records of flight segments operated by the same airline company. That is to say, using the co-booking or co-flight times to measure the strength of ties between group members will inevitably cause bias. This situation will be exacerbated by the existence of passive ties (see subsection Active Ties and Passive Ties for the concepts of passive ties) between group members inferred from large travel groups which have multiple flight segments or co-booking times.

Many factors, such as the ticket prices and travel schedules, might make a whole journey of a passenger group be related with multiple PNRs or make a PNR contain records of multiple journeys. To solve the problem, we propose a concept of **Co-Journey** behavior to cover all the group behaviors which can be observed in systems within a complete co-travel journey.

**Def. 3.** Given a passenger pair $(u, v)$, we define a **Co-Journey** event $Co\text{-}JNY(u, v) = \{Co\text{-}SFPG(u, v, i)\}$, where $i$ is one of the SFPGs in which $u$ and $v$ are together included during the whole co-travel journey. Correspondingly, we use $\#Co\text{-}JNY_{uv}$ to denote the co-journey times between the passengers $u$ and $v$.

In our experiments, we compare the statistics of civil aviation co-travel networks respectively using $\#Co\text{-}JNY_{uv}$, $\#Co\text{-}PNR_{uv}$ and $\#Co\text{-}SFPG_{uv}$ to measure the strength of ties.

### 4.3 Calculating Co-Journey Times

Based on the previous definitions, we propose an incremental algorithm to calculate the co-journey times of edges in co-travel networks. Our algorithm is a two-step state machine which can process chronologically ordered SFPGs incrementally. For the first step, we model several co-travel states for human co-travel behaviors. We design a co-journey events discoverer for passenger groups according to their transition of co-travel states. In the second step, we design a model to count co-journey times.

#### 4.3.1 Step 1: Discovering New Co-Journey Events

For a passenger pair $(u, v)$, we define two states for their daily living status during a period of time, i.e., the ***Inside-of-Co-Travel*** state and the ***Outside-of-Co-Travel*** state. As mentioned in subsection Business Scenarios of Passive Ties, passive ties are more likely to be generated from LPGs than from SPGs, and passive ties and active ties will be processed differently at the start of calculating the co-journey times, so it is necessary for us to distinguish the ***Inside-of-Co-Travel*** states in LPGs and in SPGs. Therefore, we divide the ***Inside-of-Co-Travel*** state into two sub-states, i.e., the ***Inside-of-LPG-Co-Travel*** state and the ***Inside-of-SPG-Co-Travel*** state.

We design a state machine to discover co-journey events, in which the Co-SFPG data stream can be processed in an incremental manner. For simplicity, we use ordinal numbers to denote all kinds of conditions with regard to the underlying processing event $Co\text{-}SFPG(u, v, i)$ as follows:

① The size of passenger set of the current SFPG $i$ is equal to or larger than a given threshold $T_{size}$, i.e., $|V'_i| \geq T_{size}$;

② The extent of member overlapping between the current SFPG $i$ and the previous SFPG $i-1$, indicating their last co-flight, is equal to or larger than a given threshold $T_{overlap}$, i.e., $Overlap(i, i-1) \geq T_{overlap}$, where

$$Overlap(i, i-1) = \frac{|V'_i \cap V'_{i-1}|}{|V'_i \cup V'_{i-1}|};$$

③ The interval between the date of the current SFPG $i$ and the start date of current co-journey $k$ is equal to or larger than a given threshold $T_{interval}$, i.e., $Interval(i, k) \geq T_{interval}$. It can be seen as a timeout value used to judge if we should put an end to the current co-journey.

④ The destination of the current SFPG $I$ is the same as the place of departure of the current co-journey $k$ (i.e., the starting place of the first SFPG in $k$), which can be denoted as $GoBack(i, k)$.

The threshold $T_{size}$ is setup to judge whether a passenger group is an LPG; $T_{overlap}$ is setup to determine whether two SFPGs belong to the same passenger group; $T_{interval}$ is setup to **denote** the largest travel duration of a journey, which is related with the size of the current SFPG *i*. The selection of the values of these thresholds will be further discussed in subsection Threshold Setting. The function *GoBack*(*i*, *k*) is designed to indicate whether passengers *u* and *v* have returned to their departure place of a co-journey.

The state machine discovers co-journey events between passenger *u* and *v* from their related chronologically ordered SFPGs. The state transition diagram is shown in Figure 4. When SFPG *i* comes, the machine switches the state according to the current state and the rules as follows:

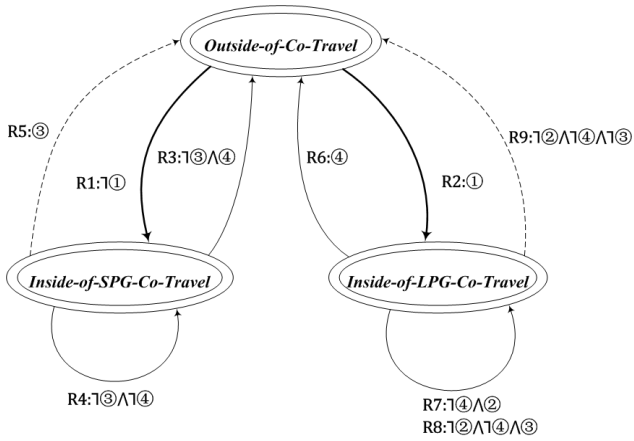

**Figure 4. State transition diagram to discover co-journey events**

**R1**: When the current state is ***Outside-of-Co-Travel*** and ⌐① happens (which means *u* and *v* appear together in a small passenger group), we make the transition
***Outside-of-Co-Travel→Inside-of-SPG-Co-Travel***, i.e., a new co-journey of *u* and *v* in an SPG begins.

**R2**: When the current state is ***Outside-of-Co-Travel*** and ① happens (which means *u* and *v* appear together in a large passenger group), we make the transition
***Outside-of-Co-Travel→Inside-of-LPG-Co-Travel***, i.e., a new co-journey of *u* and *v* in an LPG begins.

**R3**: When the current state is ***Inside-of-SPG-Co-Travel*** and ⌐③∧④ happens (which means *u* and *v* have returned to the departure place in a short time), we make the transition
***Inside-of-SPG-Co-Travel→Outside-of-Co-Travel***, i.e., their current small group co-journey is over.

**R4**: When the current state is ***Inside-of-SPG-Co-Travel*** and ⌐③∧⌐④ happens (which means *u* and *v* have not returned to the departure place in a short time), we do not make any state transition.

**R5**: When the current state is ***Inside-of-SPG-Co-Travel*** and ③ happens (which means *u* and *v* co-appear again in the same passenger group after a very long time), we first make the transition
***Inside-of-SPG-Co-Travel→Outside-of-Co-Travel***(i.e., the previous co-journey should be over) and then run the machine again with the current SFPG *i*, since it should start a new co-journey.

**R6**: When the current state is ***Inside-of-LPG-Co-Travel*** and ④ happens (which means *u* and *v* have returned to the departure place together), we make the transition
***Inside-of-LPG-Co-Travel→Outside-of-Co-Travel***, i.e., their current large group co-journey is over.

**R7**: When the current state is ***Inside-of-LPG-Co-Travel*** and ⌐④∧②happens (which means *u* and *v* have not returned to the departure place and have not changed passenger group), we think they are still in the current large group co-journey and do not make any state transition.

**R8**: When the current state is ***Inside-of-LPG-Co-Travel*** and ⌐④∧⌐②∧⌐③ happens (which means *u* and *v* have not returned to the departure place but appear together in another passenger group in a short time), we think they are still in the current co-journey and do not make any state transition.

**R9**: When the current state is ***Inside-of-LPG-Co-Travel*** and ⌐④∧⌐②∧③ happens(which means *u* and *v* have not returned to the departure place but have been in the other large passenger group after a very long time), we first make the transition
***Inside-of-LPG-Co-Travel→Outside-of-Co-Travel***(i.e., the previous large group co-journey should be over) and then run the machine again with the current SFPG *i*, since it should start a new co-journey.

Let us briefly summarize the state machine for discovering new co-journey events as shown in Figure 4. Given a sequence of SFPGs of the passenger *u* and *v*, the rules **R1** and **R2** start a new co-journey event, then the rules **R4**, **R7** and **R8** merge related SFPGs contained within the same co-journey, and finally the rules **R3** and **R6** end the current co-journey event. Specially, the rules **R5** and **R9** are used to detect those co-journeys whose SFPG sequences are not complete. The difference between **R3** and **R6** is based on the conclusion that an LPG has higher probability to book the return tickets than an SPG deduced from Figure 3 in subsection Dataset and Basic PNR and SFPG Statistics, so we can reasonably use the *GoBack*(*i*, *k*) event to identify whether the current co-journey is over; Similarly, the difference between **R8** and **R9** is based on an assumption that an LPG usually has less probability to start a new journey in a short period of time because of its low flexibility.

### 4.3.2 Step 2: Calculating Co-Journey Times

After discovering new co-journey events, including co-journeys in small passenger group (denoted as new **SPG-Co-Journey** events) and in large passenger group (denoted as new **LPG-Co-Journey** events), the next thing is to calculate the times of co-journey events.

To process new **SPG-Co-Journey** events and new **LPG-Co-Journey** events in different ways at the start of the calculating, for each pair of passengers(*u*, *v*), we define a set of states{*Null*, *#Co-JNY$_{uv}$*=0, *#Co-JNY$_{uv}$*=1, *#Co-JNY$_{uv}$*=2, *#Co-JNY$_{uv}$*>2}. Here the *Null* state indicates that no co-journey event has ever been found between *u* and *v*; the *#Co-JNY$_{uv}$*=0 state indicates that only one **LPG-Co-Journey** has been found between *u* and *v* so far, so it is an uncertain state from which we cannot tell whether they are acquaintances or strangers; while the other two states are all certain states that we can almost certainly say *u* and *v* are acquaintances.

We design another state machine to calculate the co-journey times from a chronological sequence of co-journey events between the passenger *u* and *v*. We still use ordinal numbers to denote the conditions with regard to the underlying processing co-journey event *k*:

①' The current new co-journey event *k* occurs in an LPG (i.e., *k* is a new **LPG-Co-Journey** event);

②' The current new co-journey event *k* occurs in an SPG (i.e., *k* is a new**SPG-Co-Journey** event);

The state transition diagram is show in Figure 5. When a co-journey event *k* comes, the machine switches the state according to the current state and the rules as follows:

**R1'**: When the current state is *Null* and ①' happens (which means *u* and *v* starts a new journey together in a large passenger group), we make the transition *Null*→*#Co-JNY$_{uv}$*=0, i.e., we create a passive edge between *u* and *v* to indicate the uncertain state of their relationship.

**R2'**: When the current state is *Null* and ②' happens (which means *u* and *v* starts a new journey together in a small passenger group), we make the transition *Null*→*#Co-JNY$_{uv}$*=1, i.e., we create a new active edge between them.

**R3'**: When the current state is*#Co-JNY$_{uv}$*=0 and ①'∨②' happens (which means *u* and *v* start another journey together either in a large or small passenger group),we make the transition *#Co-JNY$_{uv}$*=0 →*#Co-JNY$_{uv}$*=2, i.e., we retroactively consider their last LPG-Co-Journey as an SPG-Co-Journey because we think they are actually acquaintances.

**R4'**: When *#Co-JNY$_{uv}$*≥1 and ①'∨②' happens, we increase *#Co-JNY$_{uv}$* by 1, i.e., when we identify two passengers are acquaintances, we will not differentiate their co-journey events being in small or large passenger groups any longer.

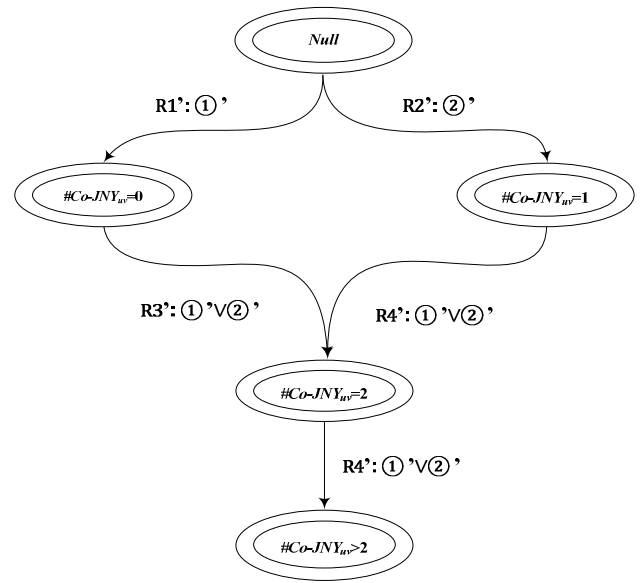

**Figure 5. State transition diagram to calculate co-journey times**

It is notable that $\#Co\text{-}JNY_{uv}=0$ does not mean that the passengers $u$ and $v$ have not travelled together ever, but only shows that we cannot determine whether there is a certain relationship between them by the existing evidence in the time window.

### 4.3.3 Threshold Setting

After the description of how to calculate Co-Journey Times, we introduce three types of thresholds as follows:

*The threshold of passenger group size*

In the state machines as described above, the different characteristics of LPGs and SPGs are very important for us to discover co-journey events and calculate the strength of ties. So we use a threshold $T_{size}$ to judge whether a passenger group is an LPG or an SPG. In this paper, we set $T_{size}$ to be 10 according to the business principle of civil aviation in China: that a passenger group containing 10 or more persons can be treated as an LPG which is able to apply for a group discount. By such a principle, some organizers sometimes intend to organize a large group containing at least 10 passengers. The rationale for setting $T_{size}=10$ is also corroborated later in this paper.

*The threshold of SFPG overlapping degree*

The threshold of SFPG overlapping extent $T_{overlap}$ is set up to judge whether two SFPGs belong to the same passenger group. To select an appropriate value of $T_{overlap}$, we test the overlapping degree of SFPGs contained in large round PNRs (i.e., the PNR that contains at least 2 SFPGs each of which has at least 10 distinct passengers which can form a round journey). We collect 992,573 PNRs of this kind and we have the distribution of numbers of PNRs over the SFPG overlapping degree as shown in Figure 6. We find that more than 97.1% of the PNRs have an SFPG overlapping degree larger than 0.9 and the degrees of 99.5% of those are larger than 0.7. Based on this fact, we set $T_{overlap}=0.7$.

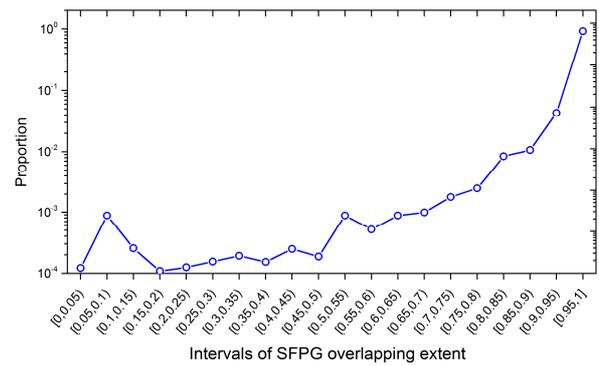

**Figure 6. Distribution of numbers of PNR over SFPG overlapping degree**

*The threshold of passenger group travel duration*

We select all PNRs each of which contains a round trip of the same passenger group from the dataset. Then we calculate the travel duration in days of each group and have average and variance of different-sizes-durations of domestic travel groups as shown in Figure 7. We found that generally, along with the increasing of the size of a group, the average of group travel duration increases and it reaches the highest point when the group size is 10 and finally keeps stable between the ranges of 5.25 - 5.5 days. However the variance of group travel duration decrease which reaches the stable minimum point when the group size is 10 or more (It is notable that 10 is the business booking threshold of large travel group.). The result shows that the thresholds for LPG and SPG should be different because of the difference of motility between them.

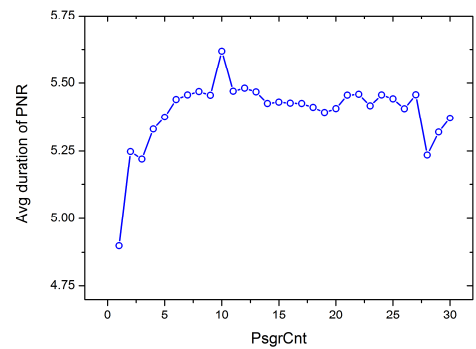

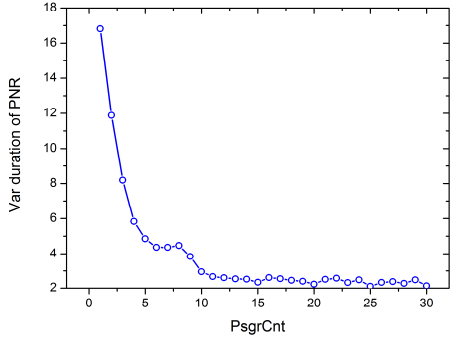

**Figure 7.** Average (a) travel durations and (b) variances of different passenger group sizes

To set an appropriate threshold for large passenger group travel duration, we let *LRPNR* be the number of large round PNRs (which means that the group journey of all large SPFGs in such a PNR is a round one.) and $LRPNR(T_t)$ be the number of round large PNRs whose duration is larger than $T_t$. Let $P_1(T_t) = LRPNR(T_t) / LRPNR$, which denotes the ratio of the number of large round PNRs whose duration are larger than $T_t$ to the total number of the large round PNRs in the sample. We got the result as shown in Figure 8. We can see that the number of large PNRs travelling for more than 15 days only make up a small proportion to the total large PNRs.

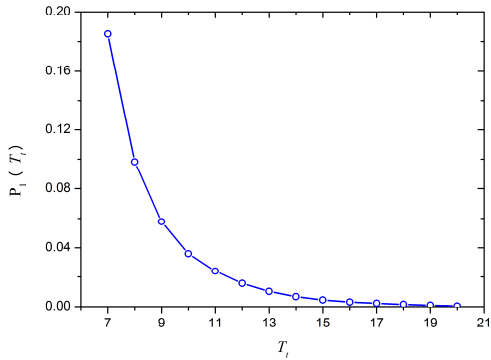

**Figure 8.** $P_1(T_t) = LRPNR(T_t) / LRPNR$

We also count the ratio of *LRPNR* to the number of all large group PNRs. It is about 65.83%. That is to say, if we set the threshold to 15 days, for any unrounded large group *LG*, the possibility of the event *LG*'s duration is larger than $T_t$ would at most be $P_2(T_t) = 0.3417 \times P_1(T_t)$ as shown in Figure 9. And in reality, the possibility is much smaller as a round PNR consists of at least two flight segments while there are at most 34.17% of large PNRs that consist of only one segment. Many of them are actually one way or open tickets, which mean the passengers in large group may have used other transportation tools for their journey or passengers have not booked their other tickets in the same PNR. Many of them may have returned to their departure place in $T_t$ days.

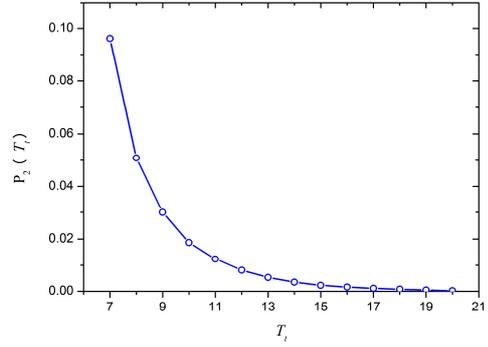

**Figure 9.** $P_2(T_t) = 0.3417 \times P_1(T_t)$

According to these facts, we set our threshold to be 15 days. One may argue why not to set a larger threshold for example 20 days. As to the error rate, according to the data, it makes no significant improvement. While, if we enlarge the time window to 20 days or more, it will leave a larger opportunity for the time window to contain two or more round journeys as the average travel duration of large groups ranges from 5.25 to 5.5 days.

For the thresholds of the travel durations of different sizes of small passenger groups, we analyze the variances of them shown in Figure 7. It shows that there exists a large difference between the variances of small passenger groups. According to this phenomenon, we set different thresholds to the durations of different sizes (e.g., 22 days to the 2-person passenger groups and 16 days to the 9-person passenger groups). Though we set larger thresholds for SPGs than the one for all LPGs, it does not mean that we cannot find multiple co-journeys within a time period of threshold length because the thresholds are timeout values used only for our machine to put an end to possible open or unfinished but timeout co-journeys. In fact we still discover new co-journey events within the given period if other evidence of ending of previous co-journey is observed.

## 4.4 Labeling Passive Ties and Active Ties

In this section, we will discuss how to label passive ties and active ties according to the co-journey times of ties. Given a time window $W$ and two passengers $u$ and $v$, there are three possible cases of the value of $\#Co\text{-}JNY_{uv}$:

**(1) $\#Co\text{-}JNY_{uv}=0$**

This means $u$ and $v$ have been observed co-traveling only once in an LPG during the time window $W$. In this case, the possible relationship between $u$ and $v$ might be strangers or acquaintances. For instance, suppose $u$ and $v$ are from the same family and joined a large tour group, then they are acquaintances but the other people in the group are likely strangers to them. Unfortunately, it is impossible for us to judge whether any two passengers in the same LPG are strangers or acquaintances without any further additional information. In desperation, we regard the passengers who appear together in an LPG for the first time as strangers, and consequently label the ties with $\#Co\text{-}JNY_{uv}=0$ as passive ties temporarily. As time goes on, if we find they co-travel again, we can immediately identify they are acquaintances with active ties.

**(2) $\#Co\text{-}JNY_{uv}=1$**

This means $u$ and $v$ have been observed co-traveling only once in an SPG during the time window $W$. In the civil aviation practice, the members in SPGs are more likely to be acquaintances than those in LPGs so we can roughly regard them as acquaintances. Thus we temporarily label the ties with $\#Co\text{-}JNY_{uv}=1$ as active ties without any further information. Once they are observed appearing in a new co-journey, their active ties are confirmed immediately.

**(3) $\#Co\text{-}JNY_{uv}\geq 2$**

This means $u$ and $v$ have been observed co-traveling twice or more during the time window $W$. In this case, we can nearly affirm they are close acquaintances and do not differentiate their large group co-journeys and small group co-journeys any more. In other words, we label all the ties with $\#Co\text{-}JNY_{uv}\geq 2$ as active ties. So we can say that $\#Co\text{-}JNY_{uv}=2$ is a sufficient condition of active ties.

## 4.5 Threshold Co-Travel Networks

To compare the statistics of co-travel networks which employ different measures of the strength of ties (i.e., $\#Co\text{-}SFPG_{uv}$, $\#Co\text{-}PNR_{uv}$ and $\#Co\text{-}JNY_{uv}$), we define three categories of co-travel networks and use a threshold to generate a family of networks with different strength of ties respectively.

**Def. 4**. We define a **Co-SFPG network** $G_s(V_s, E_s; \tau_s)$ comprising the edges $E_s$ between the passengers $V_s$, and the Co-SFPG times of each edge $(u, v)\in E_s$ should be equal to or larger than a given threshold $\tau_s$, i.e., $\#Co\text{-}SFPG_{uv}\geq\tau_s$, where $\tau_s>0$.

**Def. 5.** We define a **Co-PNR network** $G_p(V_p, E_p; \tau_p)$ comprising the edges $E_p$ between the passengers $V_p$, and the Co-PNR times of each edge $(u, v)\in E_p$ should be equal to or larger than a given threshold $\tau_p$, i.e., $\#Co\text{-}PNR_{uv}\geq\tau_p$, where $\tau_p>0$.

**Def. 6.** We define a **Co-Journey network** $G_j(V_j, E_j; \tau_j)$ comprising the edges $E_j$ between the passengers $V_j$, and the Co-Journey times of each edge $(u, v)\in E_j$ should be equal to or larger than a given threshold $\tau_j$, i.e., $\#Co\text{-}JNY_{uv}\geq\tau_j$, where $\tau_j\geq 0$.

By setting the thresholds to be different sequential integers for each category of co-travel networks, we can get three families of co-travel networks.

Given the same set of passengers $V_s=V_p=V_j$, if we do not filter out any edges, the three categories of networks $G_s(V_s, E_s; 1)$, $G_p(V_p, E_p; 1)$ and $G_j(V_j, E_j; 0)$ have the same number of edges, i.e., $|E_s|=|E_p|=|E_j|$.

In Co-SFPG networks and Co-PNR networks, a tie may be a passive tie even if it has a large $\#Co\text{-}SFPG_{uv}$ or $\#Co\text{-}PNR_{uv}$, so it is hard to give a definite threshold to filter out passive ties. For example, even if $\tau_s$ or $\tau_p$ is set to be 4 or larger, it still cannot make sure that all the passive ties in Co-SFPG or Co-PNR networks can be completely filtered out, although many active ties would inevitably be filtered out consequently. However, in Co-Journey networks, passive ties exist only in network $G_j(V_j, E_j; 0)$, thus we can easily filter them out accurately, which is very crucial for specific analyses on passive or active ties in real applications.

Let us consider a typical analysis of co-travel networks. In a Co-Journey network, we can easily and accurately pick out its backbone network components at different levels by setting a series of $\tau_j$, but it does not work in Co-SFPG or Co-PNR networks.

## 5. Network Statistics

Having defined three categories of threshold co-travel networks, now we study their structural characteristics by investigating a range of statistics. To do this, we generate networks for values of $\tau_s$ and $\tau_p$ ranging from 1..15 and $\tau_j$ ranging from 0..15. We consider the network-level features and the node-level features which are commonly studied in complex network analysis.

### 5.1 Network-level Features

At the network level, we consider the features of fractional network size, node degree distribution, fractional size of largest component, number of disconnected components and fraction of network components with different size.

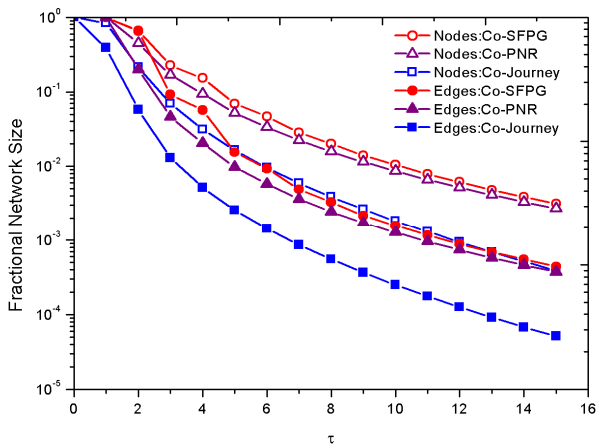

**Figure 10: Fractional network size in terms of nodes and edges over different thresholds $\tau = 2..15$, with respect to the network $G(V,Es, 1)$**

We first investigate the variation in the network size as a function of the threshold $\tau$. Figure 10 shows the number of edges and (non-isolated) nodes for each three threshold networks with the change of threshold. For instance, by increasing threshold from $\tau = 1$ (0 for $G_j$) to $\tau = 3$ or 4, the numbers of nodes and edges in the threshold network are reduced with a sharp tendency. There are two notable features in Figure 10: first, the curves of Co-Journey network are looks smoother than the other two threshold networks at the values of threshold with $\tau = 2,3,4$; and second the tendencies of the Co-SFPG and Co-PNR are similar while the Co-Journey network diminished at very different rate to the others. The explanation for these different results is as follows. The distribution of edge weights is similar between the two threshold networks of Co-SFPG and Co-PNR networks, thus the rate at which edges and nodes are removed with the increasing of $\tau$ is also similar. On the contrary, the edge weight distribution of Co-Journey network is quite different from the others to result in the difference of tendency. On the other side, the Co-Journey network contains many more "peripheral" nodes and "passive" ties at the certain low value of threshold ($\tau = 0$);however, the Co-SFPG and Co-PNR networks can only disperse such nodes and ties at different thresholds such as $\tau = 2, 3, 4$ or other large number. Thus the rate at which nodes become isolated in Co-Journey networks is initially ($\tau = 2, 3, 4$) much greater than those in Co-SFPG and Co-PNR networks with the increasing of $\tau$.

***Distributions of node degree*** in various threshold co-travel networks are shown in Figure 11. We find that the distributions in all cases exhibit power-law. It is obvious that the power-law distributions are more clear and smoother as the threshold increases, especially in Co-Journey networks. Do you remember the aforementioned 5-multiple preference? There is a similar preference in distributions of node degree with low thresholds. The explanation is that the passive ties are filtered out accurately by threshold ($\tau = 1$) and active ties lead to the more original distinct power-law distributions. We can make a conclusion that this preference is mainly caused by the passive ties after comparing the curve with $\tau = 0$ and the ones with $\tau \geqslant 0$ in Co-Journey networks. However, this preference appears not only in the threshold network with $\tau = 1$ but also in the ones with $\tau = 2, 3, 4$ or even more for Co-SFPG networks and Co-PNR networks, indicating that it is hard to filter the passive ties by a certain threshold in these networks. It also proves the rationality and superiority of co-journey times as the measure of ties strength, compared to Co-PNR and Co-SFPG times.

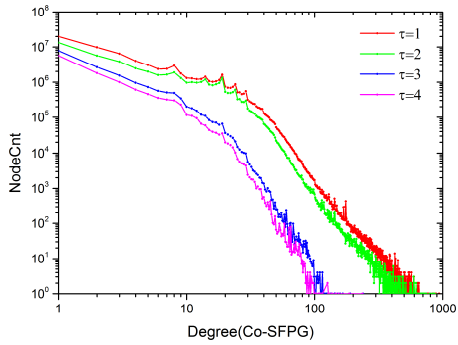

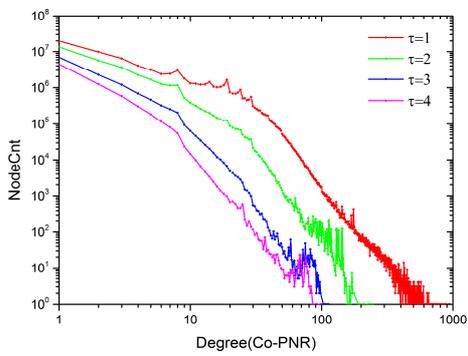

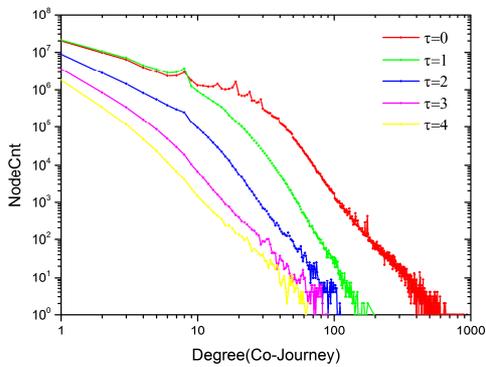

**Figure 11. Power law Distributions of Node Degree**

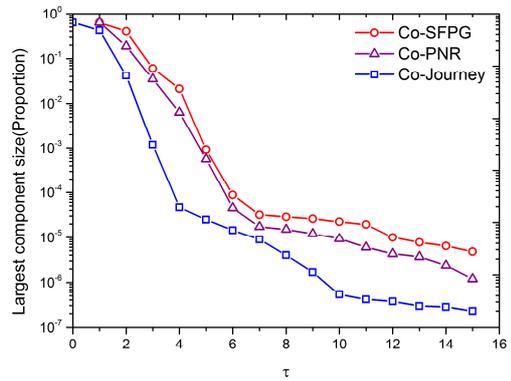

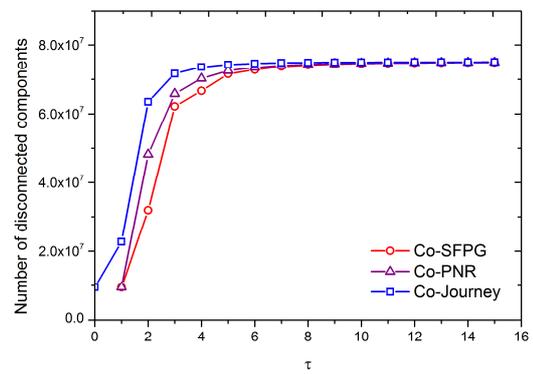

**Figure 12. (a)Changes in characteristics of the network components for each of threshold networks.(b) Number of Disconnected components of the networks of different thresholds $\tau = 0..15$**

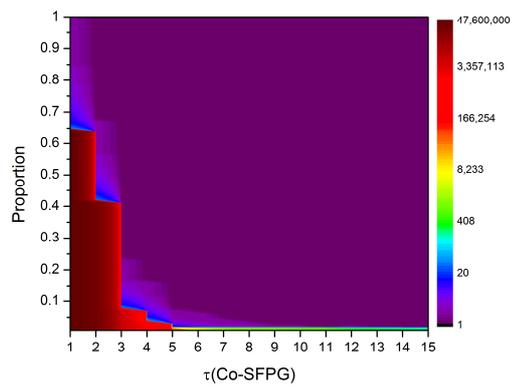

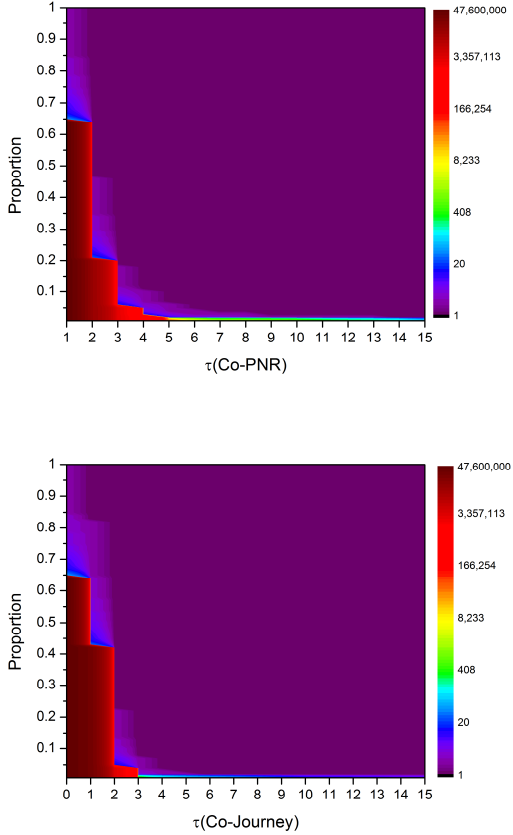

**Figure 13. Sizes of different network components as a fraction of the entire network**

Figure 12-13 present some statistics of the sizes and numbers of connected components as a function of the threshold in each network from different perspectives. Figure 12(a) shows a dramatic drop in the fractional size of the largest component in each threshold networks and Figure 12(b) shows a correspondingly dramatic increase in the number of disconnected components. In Figure 13, we can observe how the sizes of the different network components change as a fraction of the entire network in each threshold network. At τ=1(τ=0 for $G_j$), the majority of the nodes are in the largest components (size ~64%). For Co-Journey network, at around τ=5, the decomposing of connected components is almost over and for Co-SFPG and Co-PNR network, the rate of decomposing looks a little slow; the explanation is the same as the one for Figure 11.

## 5.2 Node-level Features

We consider a selected series of features at the node level. e.g., degree, two-hop neighborhood, normalized clustering coefficient, and the number of ego components. We first briefly review the definitions of these features and then present the results for all threshold networks.

*Node Degree*. The degree of a node is defined as the total number of neighbors, or immediate contacts, given by the set $\Gamma_i = \{u_j: e_{ij} \in E\}$, for individual $u_j \in V$, $k_i = \|\Gamma_i\|$.

*Size of two-hop Neighborhood*. Size of two-hop neighborhood $k_i^{(2)}$ of a node $i$ is the count of all of the node's neighbors plus all of the node's neighbor's neighbors.

*Normalized Clustering Coefficient*. The clustering coefficient of a node is a standard notion of local density(i.e., "the average percentage that two of my neighbors are neighbors of each other"), given by

$$c_i = \frac{|e_{jm}|}{C_{k_i}^2} = \frac{2|e_{jm}|}{k_i(k_i - 1)}$$

Where $e_{jm}$ are the edge between $u_j$, $u_m \in \Gamma_i$ and $\Gamma_i$ is the neighbors of *i*. The Normalized Clustering Coefficient of a node is the ratio of the clustering coefficient and the graph density:

$$C_i = \frac{c_i}{k_i/(N-1)}$$

where *N* is the number of nodes in the graph.

*Ego components*: The ego components is a count of the number of connected components that remain in its ego-network when the focal node and its incident edges are removed.

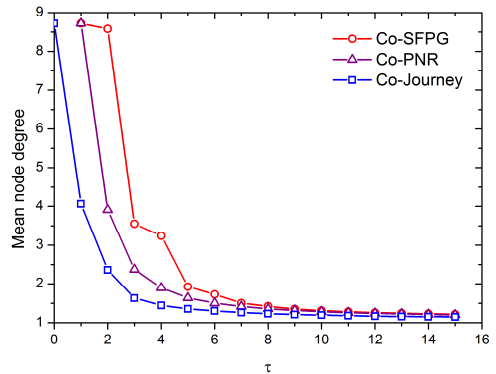

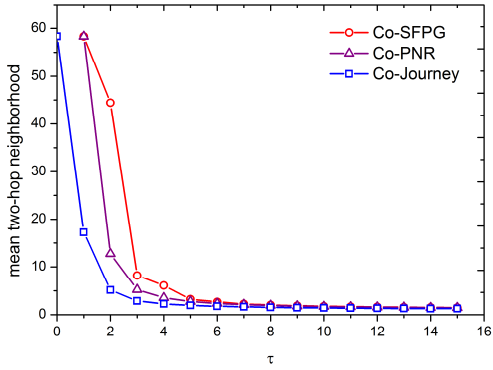

**Figure 14(a). Mean Node Degree**
**Figure 14(b). Mean Two-Hop Neighborhood**

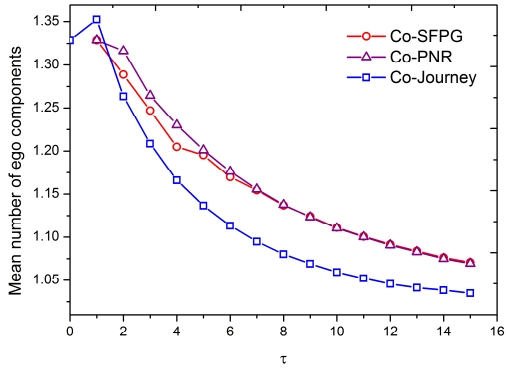

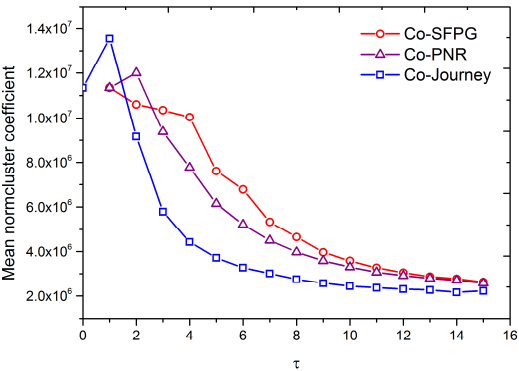

**Figure 14(c). Mean Number of Ego Components**
**Figure 14(d). Normalized Clustering Coefficient**

We study these features for the family of networks $G_s(V_s, E_s, \tau)$, $G_p(V_p, E_p, \tau)$ and $G_j(V_j, E_j, \tau)$, where $\tau$ varies between 1 (0 for $G_j$) to 15. Figure 14(a-d) shows the values of these features averaged over the population of non-isolated nodes.

The values of *Node Degree* and *Size of two-hop Neighborhood* are necessarily monotonically decreasing because increasing $\tau$ is certain to delete edges, which means every node's degree can only decrease. As can be seen in Figure 14, the average degree and two-hop neighborhood of the nodes decrease very sharply in all threshold networks, suggesting that the networks are quite sparse as a result of the cost of traveling by air.

We then consider the *Ego components*. The average number of ego components in co-travel networks is smaller than the email network [13], suggesting that in co-travel networks the average number of social circles related with a passenger is small(i.e., in real world people usually prefer to travel with those who come from very close social circles such as families or colleagues rather than the unacquainted circles.).On the other hand, Figure 14(c) indicates that the overall node trends are monotonically decreasing (where, in contrast, there is a slight increase for low values of $\tau$). The explanation for these trends appears to be that the graph comprises a number of dense clusters for low $\tau$, between which nodes can act as bridges. As we increase the threshold, however, the bridges between these clusters are preferentially severed, and aftermost ties of "this kind of bridge" are filtered out by a certain value of threshold, the other kind of bridge within clusters begin to be pruned away successively, suggesting that bridging edges outside the clusters are not as strong as those within clusters.

The changes in the measure of *Normalized Clustering Coefficient* provide further support for these hypotheses. As shown in Figure 14(d), we can see that it shows a similar variation with $\tau$ to *Ego components*. For Co-Journey networks and Co-PNR networks, as weaker, less embedded ties outside the clusters are pruned away, the clustering coefficient would increase with low value of $\tau$. And then the clustering coefficient decreases with increasing $\tau$ ($\tau \geqslant 2$),because the clusters in the networks may be mostly made up of weak ties. However we cannot find this trend in Co-SFPG networks. That is to say, the distribution of tie strength is different from Co-Journey networks, and we cannot separate the different kinds of ties by threshold effectively.

We also find an interesting phenomenon in each Figure of these features. That is, in the variation of $\tau$ there are some waves at even numbers of $\tau$ in Co-SFPG networks. This trend is quite similar to the trend of frequency distribution of PNRs with different numbers of flight segments, which we mentioned in the subsection Dataset and Basic PNR and SFPG Statistics. That may because by the tie strength measuring method, which is purely based on counting co-flight segments. Passive ties are mainly generated from LPGs and LPGs have a higher possibility of having even numbers of flight segments, which bring about these trends.

However, we cannot find such phenomena in Co-Journey networks (i.e., the curves of Co-Journey networks are smoother than the ones of Co-SFPG and Co-PNR networks for low values of threshold). It is mainly caused by our effective method that can detect complete journeys of any two passengers from their travel behavior sequences in the time window.

All the experimental results demonstrate the necessity of using Co-Journey times to measure the strength of co-travel ties and the effectiveness of our method to discover journeys composed of multiple segments. Based on such understanding, we can confirm that the Co-Journey networks are more suitable than Co-SFPG and Co-PNR networks for doing research.

## 6. Potential Applications

Besides the significance for scientific research, the high quality Co-Travel networks we inferred have many potential application scenarios especially in passenger service, flight route planning, market forecasting, etc. The networks have great potential to help provide better services for passengers, especially for group passengers. For example, overlapping communities [26, 27] can be detected from the inferred network, and it will help the aviation industry to identify whether a group is a business, tourist, or family group so as to provide better differential services. Existing information of ties with correct strength and types among passengers of a flight ready for checking-in may also possibly be used to help seat allocating agents to optimize the seat reserving strategy or plan.

Furthermore, typed information of large groups will be a very good data source for tourism attraction analysis or even for tourism service provider analysis, which are all very important decision support information for the flight route or product planning activities of airlines or governments. Evolving analysis to the Co-Travel networks will help government or airlines to better understand and forecast the evolving trends of market from the perspective of passenger social networks.

Certainly, there are many other potential applications scenarios, all of which would create great opportunities for the civil aviation industry.

## 7. Conclusions and Future Works

At the beginning of this paper, we introduced the concept about co-travel networks and discussed the solutions to infer high quality civil aviation Co-Travel Networks from passenger behavior data. We suggested using times of complete co-travel journey instead of Co-SFPG times or Co-PNR times to measure the strength of ties in networks. The necessity and advance of using Co-Journey times are well proved by the experimental results which also prove the effectiveness of our incremental algorithm to detect and count complete journeys of any two passengers from their travel behavior sequences.

By investigating the features of aviation co-travel networks from different perspectives, we discover several interesting indistinguishable features or phenomena in the networks, such as high sparsity and aggregation.

Finally we propose some potential applications of co-travel network research in civil aviation.

We note that, although the focus in this paper has been on inferring networks from the detailed travel record data with PNR information, social networks of passengers may be constructed or complemented from other kinds of observable data too. Other behaviors may also be additionally used to identify passenger groups that are not in the same PNR if we can observe these behaviors. We can establish a more complete co-travel network as more information about passenger behaviors becomes available. That is to say, although we have not used other information such as the co-flight, close-seat, close-check-in data to construct the network, these data can be used to complete and strengthen the co-travel networks.

Our future work will focus on the evolution analysis of co-travel networks so as to investigate the evolving characteristic of co-travel networks.

## 8. Acknowledgements

We are grateful to the Civil Aviation Administration of China (CACC) and Travelsky Technology Limited for their funding and data gathering. Thanks also goes to many students that helped finishing our experiments or contributed some ideas, they are Rui Tang, Yuanwei Zhou, Fuyuan Liu, Feng Wang, Ming Han, Rui Jiang, Zhiwei Wang, and Kunkun Wang. Thanks also go to the Intelligent Systems Lab of University of Bristol that provided good environment for the first author to partially finish this work in UK.